\documentclass[twocolumn,prl,tighten,floatfix,showpacs]{revtex4}
\usepackage{graphicx}

\def\<{\langle}
\def\>{\rangle}

\def\be{\begin{equation}}
\def\ee{\end{equation}}
\def\Sent{S_{\rm ent}}

\begin{document}
\preprint{cond-mat} \title{Zero dimensional area law in a gapless fermion system}

\author{G. C. Levine and D. J. Miller}

\address{Department of Physics and Astronomy, Hofstra University,
Hempstead, NY 11549}

\date{\today}

\begin{abstract}
The entanglement entropy of a gapless fermion subsystem coupled to a gapless bulk by a "weak link" is considered. It is demonstrated numerically that each independent weak link contributes an entropy proportional to $\ln{L}$, where $L$ is linear dimension of the subsystem. 
\end{abstract}

\maketitle
\section{Introduction} Striking connections have recently emerged between quantum entanglement and several distinct fields of physics: black hole quantum mechanics, quantum phase transitions in condensed matter, topologically ordered phases in quantum field theories (QFT) and conformal field theory (CFT). The central quantity common to these studies is the {\sl entanglement entropy}, computed for a finite subregion of a QFT or many body system. If, in a QFT in $d$ spatial dimensions, a distinguished region $A$ of volume $L^{d}$ is formed,  it follows that the degrees of freedom which reside exclusively in the region $A$ will appear to be in a mixed state.  The degree of mixing may be characterized by the von Neumann entropy, $S = -{\rm tr}\rho \ln{\rho}$, where  the reduced density matrix $\rho = {\rm tr}_{\notin A}|0\>\<0|$ has been formed by tracing over the degrees of freedom of the ground state, $|0\>$, exterior to the region $A$.

Entanglement entropy was first investigated in the context of black hole quantum mechanics and Hawking-Bekenstein entropy \cite{BirrellDavies}, where it was found that entanglement entropy is not a conventional extensive quantity but, rather,  is proportional to the area of the bounding surface, $S \propto L^{d-1}$ \cite{Srednicki}.  This highly suggestive result (known as the area law) is believed to bear some relation to holographic principle proposals \cite{Bousso}.  It has also recently been discovered that the subleading term in the entanglement entropy carries information about topologically ordered phases occurring in topologic QFTs \cite{top_ent}.  As pointed out in references \cite{Vidal,Cardy_rev,Korepin,QPT}, the divergence of entanglement entropy at quantum critical points may also be exploited to identify quantum critical phenomena and quantum phase transitions.  In connection with entanglement entropy,  $1 + 1$-dimensional CFTs have received special attention in the past several years.  CFTs---which describe critical spin chains, Luttinger liquids and other massless theories---have pointlike  bounding surfaces;  remarkably, the entanglement entropy was shown to depend universally upon the central charge of the theory and to diverge logarithmically with the length of the subsystem \cite{Holzhey,Vidal}.  Specifically, the entropy is given by  $S =  \frac{c}{3} \ln{L/\epsilon}$ where $c$ is the central charge.

So far, studies have concentrated on entanglement entropies computed in a homogeneous system; specifically, the bipartite entropy of a $d$-dimensional distinguished region separated from the $d$-dimensional bulk by a $(d-1)$-dimensional boundary.  For gapless fermions it has been proven that the area law is anomalous and, in contrast to the area law above,
\begin{equation}
\label{area_law}
S = \left(\frac{L}{\epsilon}\right)^{d-1}\ln{\frac{L}{\epsilon}}
\end{equation}
where $\epsilon$ is a spatial cut-off \cite{ferm1,ferm2,ferm3,ferm4}. One might consider a generalization in which a $(d-k)$-dimensional boundary separates the distinguished region from the bulk, where the codimension of the boundary, $k$, is something other than $k=1$. To maintain the appropriate extensivity of the entanglement entropy, one might expect $S \sim L^{d-k}\ln{L}$, although such a property has not been established.  In this paper we study the simplest interesting subcase: a zero dimensional boundary---or "weak link"---connecting two $d$-dimensional regions (see fig. 1). 
\begin{figure}[ht]
\includegraphics[height=6cm]{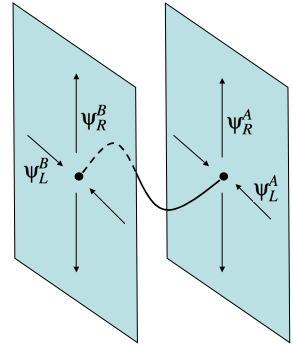}
\caption{Two $d$-dimensional regions connected by a "weak link".}
\end{figure}
For $d$-dimensional gapless fermions, this geometry may be treated by bosonization and mapped to the problem of entanglement in a $1$-dimensional subsystem where the entropy is known. The treatment of the weak link is similar to the x-ray edge or Kondo problem, where the impurity interacts with the $s$-wave sector of the bulk fermion system \cite{GNT}. The weak link leads to the entanglement of effectively two (length $L$) 1-d systems and the entropy is thus proportional to $\ln{L}$. Each independent weak link contributes an entropy proportional to $\ln{L}$, consistent with the bulk area law (eq. \ref{area_law}) if one considers a large number of weak links proportional to $L^{d-1}$.  In a sense, the area law (eq. \ref{area_law}) may be anticipated by the fusion of two $d$-dimensional regions in this fashion. 

\section{Model Hamiltonian and Bosonization}
In this manuscript, we study the entanglement entropy of a 2-d or 3-d noninteracting fermion system ($A$) connected by a weak link to a second identical system ($B$). The model hamiltonian is:
\begin{eqnarray}
\label{ham}
H &=& -t\sum_{\< i,j \> \alpha=A,B}{(c^{\dagger}_{i,\alpha}c_{j,\alpha} + c^{\dagger}_{j,\alpha}c_{i,\alpha}})\\
\nonumber&-& y (c^{\dagger}_{0,A}c_{0,B} + c^{\dagger}_{0,B}c_{0,A})
\end{eqnarray}
where $i$ and $j$ are two dimensional site indices and $A$ and $B$ denote the two identical systems coupled through a hopping amplitude, $y$. For the numerical part of this study, each subsystem is taken to be a periodic square or cubic lattice of linear size $L$.  Figure 2 shows a sample calculation of the energy eigenvalues of the hamiltonian (\ref{ham}). 
\begin{figure}[ht]
\includegraphics[height=5cm]{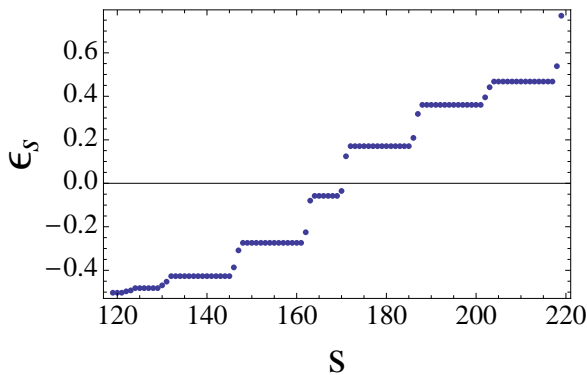}
\caption{Eigenvalues of hamiltonian (\ref{ham}) for an $L = 13$ 2-d lattice.}
\end{figure}
Unlike the comparable coupled 1-d systems for which entanglement entropy has been studied in refs. \cite{levine_imp,peschel_imp1,peschel_imp2}, the coupled $2$-d systems have extensive (in $L$) degeneracies. Moreover, the weak link only partially lifts these degeneracies in each shell to form two localized states.  These features will give rise to a special dependence of the entanglement entropy upon the filling fraction of the lattice. References \cite{ferm1,ferm2,ferm3,ferm4} have closely studied the role of critical points in the fermi surface in connection with the anomalous fermion area law.

We anticipate the central numerical results of the next section showing $S \propto \ln{L}$ by appealing to bosonization of the hamiltonian (\ref{ham}).  This model, as well as its $d$-dimensional counterparts, may be bosonized in a standard way, analogously to the kondo impurity model and x-ray edge problem \cite{GNT}. Referring to figure 1, the $s$-wave in/out fermion modes (denoted $R$ and $L$) may be combined into a single spinor
\begin{displaymath}
\Psi^\alpha(x) = \left\{ \begin{array}{cc}
\psi_R^\alpha(x) & x>0 \\
-\psi_L^\alpha(-x) & x<0
\end{array} \right.
\end{displaymath}
where $\alpha$ refers to subsystem $A$ or $B$. Since the fermion spinor now obeys periodic boundary conditions, $\Psi^\alpha(L) = \Psi^\alpha(-L)$, the model may be bosonized in the standard way,
\begin{equation}
\Psi^\alpha(x) \sim e^{i\sqrt{4\pi}\phi^\alpha(x)}
\end{equation}
where $\phi^\alpha(x)$ is a right-moving boson field on the whole interval $[-L,L]$. The hamiltonian written in chiral form is now
\begin{eqnarray}
\label{action}
H &=& \int_{-L}^L{[ (\partial_x \phi^A(x))^2 + (\partial_x \phi^B(x))^2]dx}\\
\nonumber &-& y \cos{\sqrt{4\pi}(\phi^A(0)-\phi^B(0))} 
\end{eqnarray}
This model is equivalent to the model studied analytically at weak coupling in reference \cite{levine_imp} and numerically for arbitrary coupling in reference \cite{peschel_imp1}.  Although these studies reach similar conclusions, reference \cite{peschel_imp1} definitively shows that when an impurity is introduced into a 1-d system of gapless fermions, the entanglement entropy of a subsystem of length $L$, in which the impurity lies on the boundary, is proportional to $\ln{L}$ with a nonuniversal prefactor  depending upon the impurity coupling. From the above reduction to a 1-d problem, it is plausible that each sufficiently separated weak link connecting the two 2-d subsystems would contribute an entropy proportional to $\ln{L}$. We expect entanglement to behave as
\begin{equation}
\label{zero_dim_area_law} 
S = a(y) N \ln{L} + b
\end{equation}
where the prefactor $a(y)$ depends upon the weak link coupling constant, $y$,  $N$ is the number of independent weak links and $b$ is a constant.
\section{Computation of entropy}
We compute the entanglement entropy of subsystem $A$ following the method introduced by Peschel in ref. \cite{peschel_corr_fn}. The entropy may be computed from the eigenvalues of the ground state free fermion correlation matrix
\begin{equation}
\label{corr}
C_{x,y} \equiv \<c_y c^\dagger_x \>
\end{equation}
where $x$ and $y$ are lattice points exclusively within subsystem $A$. Denoting by $\xi_k$ the eigenvalues of $C_{x,y}$, the expression for the entanglement entropy is:
\begin{equation}
\Sent = -\sum_{k}{\left((1-\xi_k)\ln{(1-\xi_k)}+\xi_k \ln{\xi_k}\right)}
\end{equation}

Figure 3 shows an example of the correlation function eigenvalues which contribute appreciably to the entropy; almost all eigenvalues are essentially zero or one. As first pointed out by Peschel \cite{peschel_corr_fn}, the set of $\xi_k$, which may be thought of as an "effective" single particle distribution function for subsystem $A$, have the characteristics of a thermal fermi distribution at a fictitious temperature (depending upon $L$ and $y$ in the present case.) Figure 3 shows a fit to the fermi function.
\begin{figure}[ht]
\includegraphics[width=7.5cm]{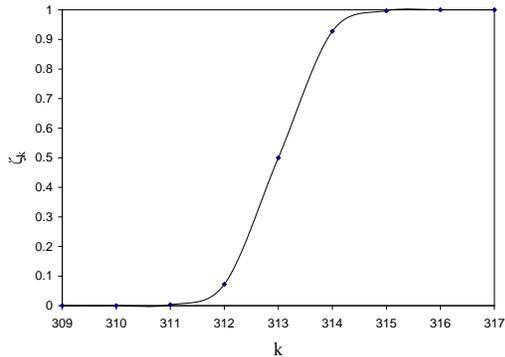}
\caption{\label{fig2} Eigenvalues of correlation function matrix (eq. \ref{corr}) for an $L =25$ lattice with $y=1$. The eigenvalues are seen to fall upon a finite temperature fermi distribution (solid line).}
\end{figure}


Figure 4 shows the entanglement entropy of a single 2-$d$ subsystem with linear size $L$ coupled to an identical subsystem. The entire coupled system has a fermion filling fraction set to $\nu = 1/2$ (that is, the number of fermions is $L^2$); for even $L$, this sets the fermi level in the middle of a degenerate shell (see fig. 2). Since the degeneracy at the fermi surface is proportional to $L$, the entanglement of these states generically leads to an entropy proportional to $L$, as seen for the even $L$ lattices in figure 4. 
\begin{figure}[ht]
\includegraphics[width=7.5cm]{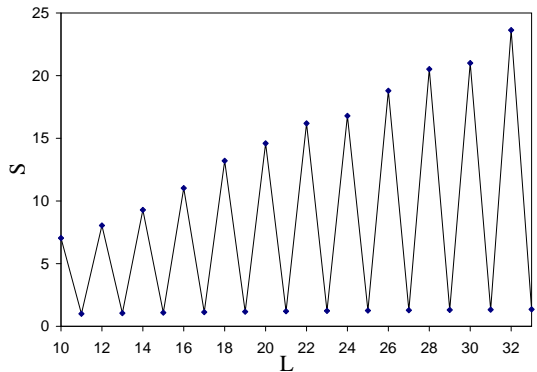}
\caption{Dependence of entanglement entropy on 2-$d$ system linear size, $L$, for weak link hopping amplitude of $y=1$.}
\end{figure}
For odd $L$ lattices at $\nu = 1/2$, this extensive degeneracy is avoided and the system better approximates the continuum limit leading to the bosonized action, eq. (\ref{action}). Returning to figure 3 showing the spectrum of eigenvalues of the correlation function, typical for an odd lattice where the extensive degeneracy is avoided, it is seen that the quasi-fermi level is at $(L^2 + 1)/2 = 313$ and only a few eigenvalues contribute significantly to the entropy.  Now, expanding the vertical scale of figure 4 and plotting only odd $L$ data versus $\ln{L}$, figure 5 clearly shows entanglement entropy proportional to $\ln{L}$, confirming eq. (\ref{zero_dim_area_law}) (for $N=1$) even for relatively small lattices. 
\begin{figure}[ht]
\includegraphics[width=7.5cm]{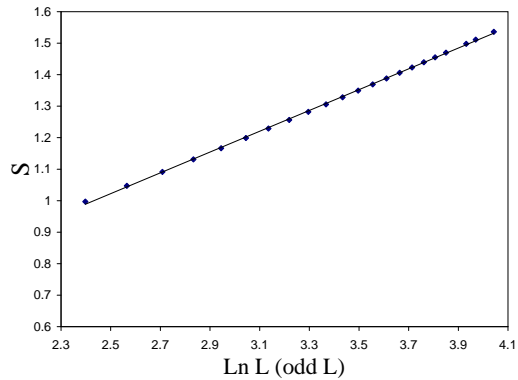}
\caption{Dependence of entanglement entropy on 2-$d$ system linear size for odd $L$ and weak link hopping amplitude of $y=1$. Linear system size in this figure ranges from $L =11$ to $L=57$.}
\end{figure}

We have also investigated the entanglement entropy in a 3-dimensional geometry. Two 3-d subsystems (obeying periodic boundary conditions) were coupled together by a single weak link as described by hamiltonian (\ref{ham}). The degeneracies in this system are more complicated than in 2-d and as a result, the degenerate shell close to zero energy is rarely completely filled for half-filled lattices ($\nu = 1/2$) with different linear sizes, $L$. Instead of fixing the value $\nu = 1/2$, the closest filling fraction to $\nu = 1/2$ was used that filled the degenerate zero energy shell.  Theses filling fractions fell within the range of $\nu = 0.497 - 0.565$ for the lattice sizes considered. Figure 6 shows the results for 3-d entanglement entropy. Entropy, $S$, is monotonically increasing with linear system size, $L$, and exhibits a periodic variation resulting from the variation in $\nu$ required to achieve the filled shell condition (see inset). Although it is impossible with our present numerics to examine the entropy over an increase in $L$ by a factor of more than approximately $3$, the behavior of entropy looks approximately proportional to $\ln{L}$ for the lattice sizes considered.
\begin{figure}[ht]
\includegraphics[width=7.0cm]{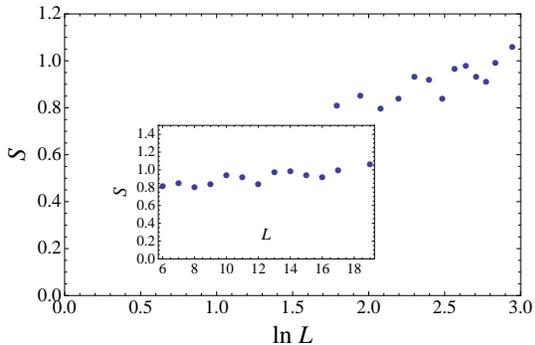}
\caption{Dependence of entanglement entropy on 3-$d$ system linear size for weak link hopping amplitude of $y=1$. Filling factor, $\nu \simeq 0.5$, is adjusted for each linear system size, $L$, to completely fill the degenerate shell closest to zero energy. Inset: entropies plotted without the logarithmic scale show periodic variation in entropy corresponding to the filled shell condition in 3-$d$. Linear system size in this figure ranges from $L =6$ to $L=19$. }
\end{figure}

Lastly, we turn to the numerical estimate of the coefficient $a(y)$, studied in 2-dimensional geometries. The dependence of the prefactor in eq.  (\ref{zero_dim_area_law}), $a(y)$, upon the weak link coupling constant, $y$, seems to be approximately quadratic for small $y$, and is shown in figure 7. Similar behavior was found for 1-D systems in reference \cite{peschel_imp1}.

\begin{figure}[ht]
\includegraphics[width=6.5cm]{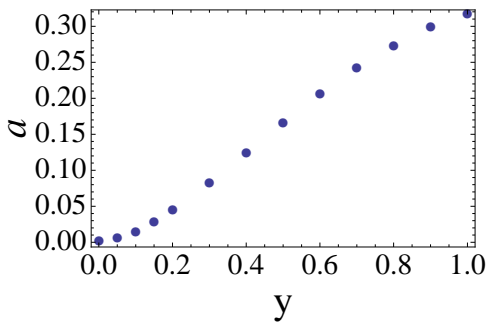}
\caption{Dependence of prefactor $a(y)$ in eq. (\ref{zero_dim_area_law}) on coupling constant $y$ for system sizes $L=15, 19$ . Small $y$ behavior is approximately quadratic.}
\end{figure}

\section{Multiple Weak Links}
We now consider the entanglement entropy of large lattices coupled by more than one weak link. The hamiltonian is the same as eq. (\ref{ham}), but modified by additional weak link interaction terms at different sites. Since each weak link lifts the degeneracy of two more states from the fermi surface, the number of fermions, $N$, must be set to be $N = L^2 - W+1$, where $W$ is the number of weak links (and $L$ is odd) to maintain the "closed shell" condition. The two-fold lifted degeneracy found for a single weak link may be seen in figure 2.
\begin{figure}[ht]
\includegraphics[width=6.5cm]{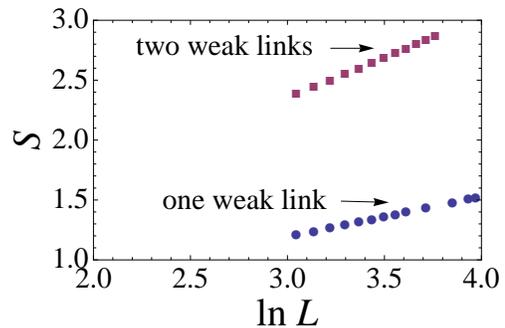}
\caption{\label{fig2} Dependence of entanglement entropy for one and two weak links on 2-$d$ system linear size for odd $L$. Consistent with zero dimensional area law, eq. (\ref{zero_dim_area_law}), slope of graph for two weak links is double that of single weak link. Weak link hopping amplitude is $y=1$.}
\end{figure}

Figure 8 shows that the entropy continues to be proportional to $\ln{L}$ for two weak links, consistent with eq. (\ref{zero_dim_area_law}). The entropies shown for two weak links in figure 8 were computed with the weak links separated maximally at $(x,y)$ locations on the lattice corresponding to $(L/4, L/4)$ and $(3L/4,3L/4)$. However, the same computation with two weak links on nearest neighbor sites generates results that are virtually indistinguishable from those of figure 8. A linear least squares fit to the $S$ versus $\ln{L}$ data in figure 8, yields a slope of $0.338$ for one weak link and double that value, $0.676$, for two weak links. We note that the intercept, $b$, in eq. (\ref{zero_dim_area_law}) was determined to be $0.168$ and $0.316$ for one and two weak links, respectively, possibly following a linear relationship $b \propto N$. Such a behavior would not be surprising---reflecting a subdominant $O(L)$ contribution to the bulk area law, eq. (\ref{area_law}). Thus the entropy accorded each weak link appears to be additive, largely independent of the spacing of the two weak links.
\begin{figure}[ht]
\includegraphics[width=6cm]{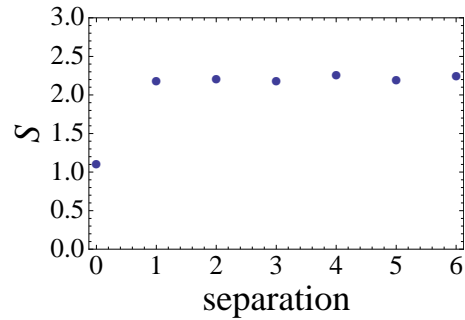}
\caption{\label{fig2} Dependence of entanglement entropy upon the separation between two weak links (with $y=1$) coupling subsystems $A$ and $B$. The zero separation data point corresponds to the entropy of a single weak link (with $y=1$) for reference.}
\end{figure}

The separation of two weak links is explored further in figure 9, where the entropy  is computed for two weak links on an $L=15$ lattice, with a separation that ranges from $0$ to $6$ lattice spacings. Other than a slight (not understood) odd/even modulation, the entropy is independent of the spacing, and its value, $S \sim 2.16-2.19$ is approximately double the entropy of a single weak link with identical coupling, $S \sim 1.09$. 

To explore the additivity of the entanglement entropy for multiple weak links, we computed entropy for up to four weak links on an $L = 25$ lattice, maintaining the maximum possible separation between them. As shown in figure 10, each additional identical weak link contributes an entropy of approximately $S \sim 1.26$, until some saturation appears for four weak links. In terms of the Fermi wavelength, $\lambda_F$, saturation begins to appear at a linear separation of approximately $12$ lattice sites, corresponding to $6\lambda_F$ for a half filled lattice. It is interesting to note that although saturation effects are seen at a spacing much larger than $\lambda_F$, two weak links separated by only a single lattice spacing appear to be independent in that they contribute entropies {\sl additively} as shown in figure 9.  

\begin{figure}[ht]
\includegraphics[width=6cm]{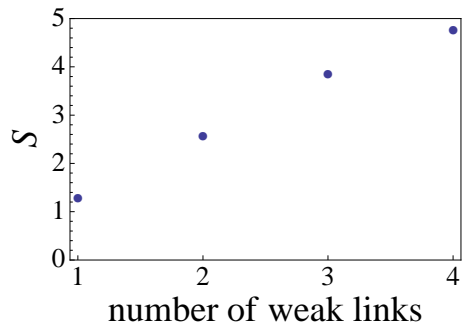}
\caption{\label{fig2} Dependence of entanglement entropy upon the number of weak links. The weak links were maximally separated in an $L =25$ lattice.}
\end{figure}

\section{Conclusion}
We have demonstrated that a weak link coupling two gapless fermion subsystems results in an entanglement entropy between the subsystems proportional to $\ln{L}$ where $L$ is the linear dimension of the subsystem (eq. \ref{zero_dim_area_law}). This result is consistent with the bulk area law for fermions (eq. \ref{area_law}) in that the entropy in a geometry containing multiple, sufficiently separated, weak links is proportional to both $\ln{L}$ and the number of weak links. The consistency may also be seen by noting that the bulk area law depends upon an arbitrary cut-off, $\epsilon$, depicted in figure 11 as a spatial cut-off taken tranverse to the boundary. Choosing a cut-off in the bulk area law (eq. \ref{area_law}), as shown in figure 11, is equivalent  to choosing a number of weak links, $N$, in the  zero dimensional area law (eq. \ref{zero_dim_area_law}) connecting the two regions satisfying,
\begin{equation}
N = \frac{1}{a(y)}\left( \frac{L}{\epsilon} \right)^{d-1}
\end{equation}

This research was supported by an award from Research Corporation CC6535. GL also wishes to thank the Kavli Institute for Theoretical Physics where this work was, in part, completed; this research was supported in part by the National Science Foundation under Grant No. PHY05-51164.

\begin{figure}[ht]
\includegraphics[height=6cm]{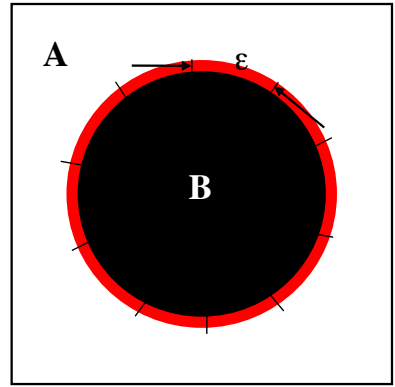}
\caption{Two $2$-dimensional regions, $A$ and $B$, connected by several weak links (shown as short radial line segments). Taking the boundary length to be $L$ and choosing a number of weak links, $N= \frac{L}{a\epsilon}$, is equivalent in the zero dimensional area law (eq. \ref{zero_dim_area_law}) to choosing a cut-off $\epsilon$ in the bulk area law (eq. \ref{area_law}). }
\end{figure}

\end{document}